\documentclass[acmlarge, screen, table, authorversion]{acmart}

\usepackage{enumitem}

\usepackage[capitalise, nameinlink]{cleveref}

\usepackage{tabularx}

\usepackage{acronym}

\apptocmd{\UrlBreaks}{\do\f\do\m}{}{}

\newcommand{\OpenClawVersion}[0]{2026.2.2-3}

\newcommand{\Repo}[0]{https://github.com/jgru/forensic-analysis-of-openclaw}
\newcommand{\Commit}[0]{98edee7}

\acrodef{LLM}{large language model}

\setcopyright{acmlicensed}
\copyrightyear{2026}
\acmYear{2026}
\acmJournal{DTRAP}
\acmDOI{10.1145/3848419}
\acmVolume{}
\acmNumber{}
\acmArticle{}
\acmMonth{10}
\makeatletter
\AtBeginDocument{%
  \let\ps@acmstandard\ps@standardpagestyle
  \def\ps@standardpagestyle{\ps@acmstandard\fancyfoot[RO,LE]{\footnotesize\@journalNameShort}}%
  \let\ps@acmfirst\ps@firstpagestyle
  \def\ps@firstpagestyle{\ps@acmfirst\fancyfoot[RO,LE]{\footnotesize\@journalNameShort}}%
  \pagestyle{standardpagestyle}}
\makeatother

\begin{document}
\pagenumbering{arabic}

\title{Foundations for Investigations of Agentic AI Systems from the Forensic Analysis of OpenClaw}

\author{Jan Gruber} 
\authornote{Corresponding author}
\orcid{0000-0003-1862-2900}
\email{jan.gruber@kit.edu}
\affiliation{%
  \institution{KASTEL Security Research Labs, Karlsruhe Institute of Technology (KIT)}
  \streetaddress{Am Fasanengarten 5}
  \city{Karlsruhe}
  \postcode{76131}
  \country{Germany}
}

\author{Jan-Niclas Hilgert}
\orcid{0009-0000-5308-5712}
\email{jan-niclas.hilgert@fkie.fraunhofer.de}
\affiliation{%
  \institution{Fraunhofer Institute for Communication, Information Processing and Ergonomics FKIE}
  \streetaddress{Zanderstr. 5}
  \city{Bonn}
  \postcode{53177}
  \country{Germany}
}

\begin{abstract}
Agentic AI systems are increasingly deployed as personal assistants and virtual employees, and are likely to become a common object of digital investigations and incident response. Yet little is known about how their internal state and actions can be reconstructed during forensic analysis, and systematic forensic approaches remain largely unexplored.
We present the first empirical forensic study of a personal AI assistant, with OpenClaw as a case study. We examine its technical design via static code analysis and apply differential forensic analysis to identify recoverable traces across stages of the agent interaction loop. We classify and correlate these traces, structuring the examination around representative investigative questions. We propose a five-plane agent artifact taxonomy that captures recurring investigative patterns; its structural completeness is assessed by cross-mapping against published artifact inventories of two related AI systems. To aid investigations, we release an open-source artifact-extraction tool with a machine-readable schema.
Finally, we highlight a foundational challenge: LLM-mediated execution adds both an abstraction layer and substantial nondeterminism to trace generation. Because tool selection and state transitions are co-determined by the model, the execution environment, and the evolving context, identical inputs need not yield identical traces.
\end{abstract}



\maketitle

\acresetall
\section{Introduction}
\label{sec:introduction}
AI agents represent a fundamental leap in artificial intelligence, transforming chatbots into autonomous systems that can reason, plan, and execute real-world workflows. The highly popular open-source software project \emph{OpenClaw} democratizes this technology, allowing non-technical users to deploy powerful agents as personal assistants without complex coding, which suggests these systems will soon become common in digital investigations.\footnote{See~\url{https://web.archive.org/web/20260219034956/https://www.wired.com/story/clawdbot-moltbot-viral-ai-assistant/}, accessed on 15.07.2026.} Despite this trailblazing trend, the forensic community currently lacks both experience with, and an understanding of, how to identify and analyze the traces %
produced by these systems. %

\subsection{Motivation}
In the past months, we have seen agentic AI rapidly evolve from tools for coding and online workflows into ``personal AI assistants'' and further into semi-autonomous entities embedded in users' daily lives.
While they resemble earlier smart-home assistants in their ability to support the user in accomplishing tasks, for instance in the form of Amazon's ecosystem~\citep{ChungPL17,CrasseltG24},
these new systems extend far beyond passive interaction: they execute multi-step workflows, integrate deeply with operating systems and online services, and are granted broad permissions over highly sensitive personal data.
Furthermore, these systems are intended for always-on assistance running 24/7
and operate as more or less autonomous entities while their actions can produce real-world effects that are not fully anticipated or even observed by the human operator. \citet{ShapiraWYSPFBLJPC26} conducted laboratory experiments and observed OpenClaw instances to show highly concerning behavior such as ``unauthorized compliance with non-owners, disclosure of sensitive information, execution of destructive system-level actions, [\dots] and partial system takeover'' while creating denial-of-service conditions, consuming resources in an uncontrolled manner. 
When investigating incidents related to AI agents, the central forensic question is not only what happened on the host, but what the agent actually did versus what the user requested, knew, or intended---an attribution problem that current forensic understanding struggles to resolve.

Consequently, a significant research gap exists in the forensic analysis of agentic assistants, rooted both in limited system understanding and the absence of dedicated investigative methodologies. Existing digital forensic approaches are ill-equipped to handle the combination of cloud-mediated reasoning, local execution, and continuous autonomy. This work addresses this gap by establishing the foundation for the forensic examination of AI assistants.

\subsection{Contributions}
\label{subsec:contributions}
To our knowledge, this work presents the first forensic analysis of a personal AI assistant, using the trailblazing open-source system \emph{OpenClaw} as a representative case study. Our contributions are threefold:

\begin{enumerate}
  \item \textbf{Systematic forensic characterization.} Based on static code analysis, differential analysis, and in-depth file system examination, we identify, classify, and correlate recoverable on-disk traces with observed agent behavior. We structure the examination around representative investigative questions and demonstrate how they can be addressed through the locally available artifacts.
  \item \textbf{Agent artifact taxonomy.} We synthesize the theoretical agent interaction loop and core architectural components with our empirical findings to propose a five-plane artifact taxonomy that provides a reusable framework for the systematic forensic analysis of agentic AI systems.
  \item \textbf{Identification of qualitatively new forensic challenges.} We demonstrate that agentic AI systems produce a fundamentally different class of forensic evidence---most notably \ac{LLM} reasoning traces with no analog in traditional software---and that their \ac{LLM}-mediated execution introduces nondeterminism and an additional layer of abstraction into trace generation, challenging long-standing forensic assumptions about reproducibility and provenance.
\end{enumerate}

To support practical analysis, we implement a prototype tool that automatically extracts and correlates local artifacts from an OpenClaw installation. Experimental data and source code are made publicly available.\footnote{See \url{\Repo}, Commit \Commit{}.}

\section{Background \& Related Work}
\label{sec:bg_related_work}
\subsection{Agentic AI and AI Agents}
\label{sec:bg_related_work:single_agent}
\emph{Agentic AI} denotes the broader architectural paradigm of systems that can autonomously collect data, reason about problems, and execute self-determined tasks with minimal human intervention~\citep{Pati25}. Such systems extend generative AI by adding planning, action, memory, and adaptation capabilities, taking initiative on behalf of users rather than merely responding to prompts~\citep{LazerAGB26,HuangH25}. Within this paradigm, an \emph{AI agent} (or \emph{single-agent system}) is a computational entity that perceives its environment, reasons about observations, and executes actions in pursuit of explicit goals with limited human oversight~\citep{SapkotaRK26,AliDC26}. OpenClaw is such a single-agent system; although it can spawn ephemeral subagents to handle scoped subtasks, these operate under the primary agent's authority and do not constitute independent autonomous entities as in multi-agent architectures.

Architecturally, AI agents are structured around an \ac{LLM} as the core reasoning component, extended with mechanisms for planning, memory access, and tool selection~\citep{HuangH25}. A control loop orchestrates their behavior~\citep{HuangH25,HuangH25b}: the agent perceives inputs, incorporates them into the \ac{LLM}'s context and memory, determines a strategy, and executes it through tool-driven actions (e.g., APIs, CLI tools, or software environments)~\citep{LazerAGB26}. It then observes the results to update its context and memory---both short-term (episodic) and persistent (retrieval-augmented)---before the next iteration~\citep{SapkotaRK26}.

\subsection{Related Work}
Prior work can be broadly grouped into two streams. First, researchers have applied machine learning and AI models to conduct or support digital forensic investigations, traditionally through narrow, task-specific methods~\citep[for an overview]{FahndrichHPRBL23} and increasingly through general-purpose \acp{LLM}~\citep{ScanlonBHHS23}, including explorations using agentic AI frameworks~\citep{WickramasekaraS24}.

Second, researchers have examined AI systems themselves as forensic objects---as subjects of investigation, instruments enabling wrongdoing, or targets of attack. Our work contributes to this latter (and comparatively sparse) stream.
\citet{SchneiderB23} examined the attribution and accountability challenges that arise when AI systems act with limited or no real-time human oversight. They highlighted the difficulty of determining whether an AI system merely executed predefined logic or exhibited behavior that could be considered independent or emergent. This distinction is central to forensic analysis of autonomous AI agents, such as OpenClaw.
In contrast to prior work that focuses on defining AI forensics, recent work has begun to operationalize AI forensics by proposing concrete sources of forensic evidence within AI systems. In this context, \citet{DragonasLN24} present a forensic analysis of the ChatGPT mobile application, demonstrating the feasibility of extracting evidence related to \ac{LLM} interaction in this mobile application. \citet{ChernyshevBD24} focused on the \ac{LLM} itself. They were the first to analyze \ac{LLM} invocation logs to reconstruct intent, usage patterns, and misuse.
As the first forensic analysis of agentic tooling, \citet{WalkerGAHB24} investigated forensic artifacts when using Microsoft's \emph{AutoGen} multi-agent \ac{LLM} framework. In their experiments, they identified logs, configuration files, agent communication traces, and execution metadata that enable reconstruction of agent behavior and task workflows. Their findings demonstrated that meaningful forensic analysis is possible in multi-agent \ac{LLM} environments and emphasized the need for specialized methodologies to support investigations involving collaborative AI systems.

While OpenClaw is an orchestration layer utilizing \acp{LLM} for reasoning, these works are related insofar as they deal with identifying internal AI artifacts---such as logs and prompts---as forensic evidence.
Our analysis of OpenClaw, however, goes beyond these settings, as OpenClaw operates as an AI assistant with substantially greater autonomy and much deeper integration into the host system, enabling persistent state, direct interaction with local resources, and self-directed task execution, which introduces a broader and more complex forensic surface.

Using the taxonomy proposed by~\citet{BehzadanB20}, the present paper is situated in the field of \emph{AI Substrate} and \emph{AI Application Forensics}, contributing to the conceptual and methodological foundations of AI forensics and providing empirical insights into the forensic analysis of agentic AI systems.

\subsection{Overview of the Functionality}
OpenClaw is a self-hosted personal AI assistant designed to translate natural-language instructions into concrete actions across connected devices and services. At a high level, the system comprises a centralized gateway and a primary agent capable of spawning subagents.

\begin{figure*}[htbp]
  \centering
  \resizebox{0.95\textwidth}{!}{
    \includegraphics{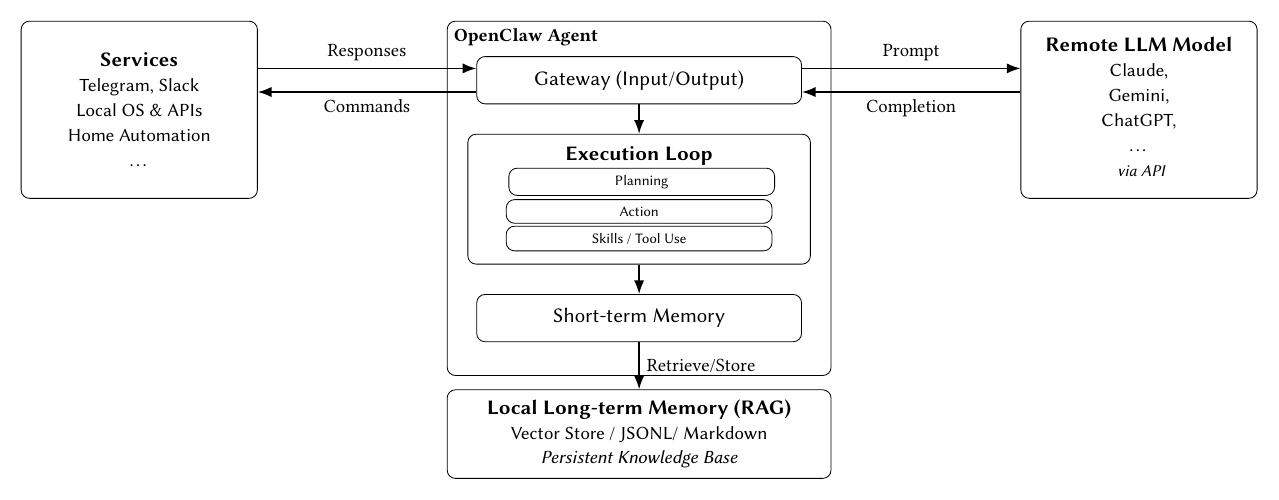}
  }
  \caption{High-level Architecture of OpenClaw.}
  \label{fig:openclaw_arch}
  \Description{Block diagram with three columns. Left: external services (Telegram, Slack, local OS and APIs, home automation) exchange commands and events with the agent. Center: the OpenClaw agent, consisting of a Gateway for input and output, an execution loop with planning, action, and skills/tool use, and a short-term memory that retrieves from and stores to a local long-term memory for retrieval-augmented generation, built on a vector store, JSONL, and Markdown files as persistent knowledge base. Right: a remote LLM accessed via API receives prompts from and returns completions to the Gateway.}
\end{figure*}

OpenClaw enables continuous interaction with users through existing messaging applications as well as dedicated companion apps for macOS, iOS, and Android. Functionally, OpenClaw combines LLM-based reasoning with device-level execution implemented through extensible skills. These skills enable operations such as sending emails, controlling a web browser, managing calendar appointments, and interacting with the local file system. The integration of persistent memory, natural-language reasoning, direct system interaction, and subagent management distinguishes OpenClaw as an agentic system. Its high-level architecture is visualized in~\cref{fig:openclaw_arch}.

From a more technical perspective, OpenClaw is designed as a modular, local-first platform to act as an orchestration layer around an \ac{LLM} at its core, services through APIs, and data stores (such as file systems). The system is implemented primarily in TypeScript and can be run via JavaScript runtime environments \texttt{Node.js} and \texttt{bun}.

\subsection{Gateway Service}
At the heart of the architecture lies the \emph{gateway}, a Node.js-based service acting as the always-on and remotely accessible central process. It employs an event-driven architecture to handle real-time communication, utilizing WebSockets for client connectivity and internal event emitters for subsystem coordination.
The Gateway is responsible for:
\begin{itemize}
[  leftmargin=1.5em,
    itemsep=-1ex,
    topsep=1ex,
    partopsep=0ex,
    parsep=1ex,
    labelsep=0.5em,]
  \item \textbf{Lifecycle Management:} Orchestrating the initialization, execution, and termination of sessions.
  \item \textbf{Routing \& Dispatch:} Multiplexing incoming messages from external channels (e.g., Telegram, WhatsApp, Slack) to the appropriate agent contexts.
  \item \textbf{Tool Execution:} Providing a runtime environment where agents can invoke local tools, ranging from executing Bash commands to browser automation.
\end{itemize}
Hence, it is the central component of the event/control-plane.

\subsection{Pluggable Communication Channels}
OpenClaw abstracts external communication platforms through a modular plugin system. Each channel (e.g., Telegram, WhatsApp via \textit{Baileys}, Discord, Slack) operates as an independent module that normalizes incoming events into a standard internal schema.
The architecture decouples the transport layer from the agentic logic. A ``Dock'' abstraction manages the connection state (e.g., WebSocket keep-alives or webhook listeners), ensuring that agent cognition remains agnostic to the specific medium of communication.

\subsection{Memory and Persistence Layer}
An important feature of OpenClaw is its reliance on local persistence for context management. The memory architecture is built upon plain text files and an SQLite database, enhanced with the \texttt{sqlite-vec} extension to support vector embeddings.
The system maintains several Markdown files where the agent stores important facts to load as context and a \texttt{memory.sqlite} containing chunked text and associated vector embeddings in a \texttt{chunks\_vec\_*} table. This allows agents to perform semantic search and retrieval-augmented generation.

\subsection{Security and Integrity}
To provide a secure environment, OpenClaw relies on the concepts of permissions, containment, and human oversight to mitigate the risk of an autonomous agent executing destructive commands.
To do so, the architecture supports a strict \textit{Policy Enforcement Model}. Sensitive operations, particularly those involving shell execution, can potentially be sandboxed using Docker containers with minimal privileges (read-only root, network isolation). Additionally, an \textit{Execution Approval Manager} gates critical actions, ensuring a ``Human-in-the-Loop'' workflow by holding high-risk tool calls in a pending state until explicit authorization is granted.

\section{Methodology}
\label{sec:methodology}
\begin{figure*}[htbp]
\centering
  \resizebox{0.9\columnwidth}{!}{
  \includegraphics{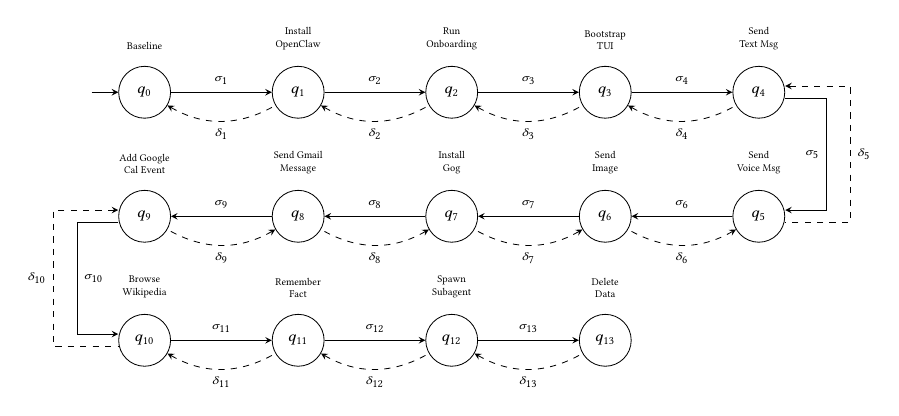}
  }
  \caption{Data generation workflow. Actions $\sigma_i$ transition the system states $q_i$, enabling differential analysis $\delta_i$ to isolate artifacts.}
  \Description{State-machine diagram laid out in three rows: fourteen states, q0 to q13, are connected by solid arrows labeled with actions sigma-1 to sigma-13. Each state is annotated with the action that produced it, such as baseline creation, installing OpenClaw, onboarding, bootstrapping the TUI, sending text, voice, image, and Gmail messages, adding a Google calendar event, remembering a fact, spawning a subagent, and deleting data. Dashed arcs labeled delta-i connect consecutive states, indicating the differential comparison of the corresponding disk images.}
  \label{fig:states_actions}
\end{figure*}
\subsection{Experimental Design}

The experiments were conducted in a virtualized environment managed by Oracle VirtualBox, the guest virtual machine operated on \emph{Debian GNU/Linux 13}. Within this virtualized environment, \emph{OpenClaw} in version~\texttt{\OpenClawVersion}\footnote{See \url{https://www.npmjs.com/package/openclaw/v/2026.2.2-3}, accessed on 15.07.2026.} was installed and operated to generate the requisite test data. We chose Linux as the experimental platform because OpenClaw is commonly run on cloud VMs and the recommended way to install the software on Windows is using WSL2, making Debian-based Linux a representative target.

Following the methodology of~\citet{KalberDF13}, we systematically generated test data by executing a series of actions ($\sigma_i \in \Sigma$) that encompass installation, setup, and various interactions with the agent. Each action transitions the system to a distinct state ($q_i$), which is captured as a disk image, starting from a baseline $q_0$. \cref{fig:states_actions} illustrates this process, enabling the isolation of artifacts tied to specific agent functionalities.

\subsection{Analysis Procedure}
Our analysis combined static code analysis, differential analysis, and file inspection to locate and interpret relevant traces. Differential forensic analysis is a well-established method~\citep{GarfinkelNY12,KalberDF13} for systematically cataloging application-specific artifacts under controlled conditions. Crucially, the artifact catalog produced by this approach is not bound to the laboratory setting in which it was derived: once the locations, formats, and semantics of relevant traces are known, practitioners can locate and interpret them.
First, we performed static code analysis of the source code. This step provided an overview of the application's architecture, its components and their main functionality, and potential data storage locations (such as database schemas or configuration files) before generating any user data.

\sloppypar Next, we applied the differential forensic analysis~\citep{GarfinkelNY12}. This allowed us to isolate the specific effects of an action by comparing the differences between two sequential images. We utilized The Sleuth Kit in conjunction with the \texttt{DFXML} library~\citep{Garfinkel12} and the Python script \texttt{idifference2.py}~\citep{Garfinkel09}, to programmatically identify changes \(\delta_i\) in the file system induced by the action \(\sigma_i\) in question by comparing two consecutive state images \(\langle q_i, q_{i+1}\rangle\). The tool detects changes across five specific categories: file or directory creation, deletion, renaming, content modification, and timestamp updates. To focus the investigation, we applied a filter to exclude system background noise unrelated to the executed actions.

Finally, the remaining candidates of file system changes were subsequently verified and inspected in-depth. We parsed the relevant files---including plain text files, SQLite databases, JSONL files, and others---using standard forensic tooling and custom scripts. The data recovered from these files were then compared against the known inputs from the test data generation phase to confirm the artifacts' content.

\section{Forensic Analysis of OpenClaw}
\label{sec:analysis}
We now present the results of our analysis. We begin with an overview of the main artifact locations, then structure our findings around the core components of the agent architecture described in \cref{sec:bg_related_work:single_agent}: the agent's cognitive engine and configured identity, its accumulated knowledge, its communication interfaces, and its executed actions. This organization reflects the modular design of agentic AI systems, where each architectural component produces forensically distinct artifacts that require different analytical approaches. For each component, we address representative investigative questions and introduce the relevant artifacts in detail.

For reproducibility and documentation, we provide the description of the experiments in literate programming style, experimental data, and a tool supporting the analysis as supplementary material~(\cref{appendix:supplementary_material}).\footnote{See~\url{\Repo}.}
There, we documented our main findings in the artifacts YAML format to feed machine-readable knowledge bases, like the \emph{Digital Forensics Artifacts Repository}~\citep{MetzArtifacts}.

\subsection{Artifact Overview}
\label{subsec:artifact_overview}
While standard Linux OS artifacts are well understood and should of course be included in an investigation, we focus in the following on artifacts specific to OpenClaw. In a standard installation, the vast majority of data linked to the AI agent resides in a hidden directory named \texttt{.openclaw} in the user's home directory. \Cref{tab:artifact_overview} summarizes the primary artifact locations. All paths are relative to this directory unless stated otherwise. Ephemeral runtime logs are written outside this directory to \texttt{/tmp/openclaw/*.log}.

\DeclareRobustCommand{\filledbullet}{\includegraphics[height=2mm]{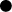}}
\DeclareRobustCommand{\halfbullet}{\includegraphics[height=2mm]{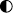}}

\begin{table*}[!t]
  \centering
  \caption{Overview of OpenClaw artifact locations and their forensic relevance. Phase columns indicate relevance to: \textbf{Char.}\,=\,Agent Characterization, \textbf{Know.}\,=\,Knowledge Reconstruction, \textbf{Inter.}\,=\,Interaction Tracing, \textbf{Act.}\,=\,Action Reconstruction. Relevance levels: \filledbullet\,=\,primary evidence, \halfbullet\,=\,secondary evidence, blank\,=\,not relevant.}
  \label{tab:artifact_overview}
  \scriptsize
  \begin{tabularx}{\textwidth}{@{} l X cccc @{}}
    \toprule
    \textbf{Path} & \textbf{Contains} & \textbf{Char.} & \textbf{Know.} & \textbf{Inter.} & \textbf{Act.} \\
    \midrule
    \texttt{openclaw.json} & Agent, model, channel, and permission settings & \filledbullet & & \filledbullet & \\
    \texttt{openclaw.json.bak*} & Historical configuration snapshots & \filledbullet & & \halfbullet & \\
    \texttt{credentials/} & Channel authentication (Telegram, WhatsApp) & \halfbullet & & \filledbullet & \\
    \texttt{agents/\{id\}/agent/auth-profiles.json} & Provider credentials (OAuth tokens, API keys) & \filledbullet & & & \\
    \texttt{devices/}, \texttt{identity/} & Paired companion devices and keys & & & \filledbullet & \\
    \texttt{workspace/*.md} & Agent persona, user profile, tool definitions, curated memory & \filledbullet & \filledbullet & & \\
    \texttt{workspace/skills/} & Custom agent skill definitions & \filledbullet & \halfbullet & & \\
    \texttt{workspace/memory/} & Daily memory logs (created on demand) & & \filledbullet & \halfbullet & \\
    \texttt{memory/\{agent\_id\}.sqlite} & Semantic search embeddings & & \filledbullet & & \\
    \texttt{agents/\{id\}/sessions/sessions.json} & Session metadata and mappings & \filledbullet & \halfbullet & \filledbullet & \halfbullet \\
    \texttt{agents/\{id\}/sessions/*.jsonl} & Full conversation and tool execution history & \halfbullet & \filledbullet & \filledbullet & \filledbullet \\
    \texttt{media/inbound/} & User-uploaded files (images, voice notes) & & \halfbullet & \filledbullet & \\
    \texttt{cron/jobs.json} & Scheduled autonomous task definitions & \filledbullet & & & \filledbullet \\
    \texttt{cron/runs/\{jobId\}.jsonl} & Execution logs for scheduled tasks & & & & \filledbullet \\
    \texttt{subagents/runs.json} & Delegated task tracking & & & & \filledbullet \\
    \texttt{/tmp/openclaw/*.log} & Runtime events & \halfbullet & & \halfbullet & \filledbullet \\
    \bottomrule
  \end{tabularx}
\end{table*}

\subsubsection{Session Transcripts}
\label{subsubsec:sessions}
Session transcripts are the single richest forensic artifact in the system, as they record not only communication but also reasoning traces, tool invocations, and context state. Because they span all aspects of the agent's operation, we introduce them here before turning to the component-specific analysis.

OpenClaw organizes conversations using a two-level hierarchy.
A \emph{session} represents a long-lived conversation context and is identified by two complementary keys:
(i) a human-readable \texttt{sessionKey} (e.g., \texttt{agent:main:main}) used for routing, and
(ii) a UUID-based \texttt{sessionId} that uniquely identifies a session instance.
Sessions persist across interactions and record metadata such as the working directory, model/provider overrides, the configured thinking level, and the path to the transcript file.
A session transitions through the states \texttt{idle} (awaiting input), \texttt{processing} (actively executing), and \texttt{waiting} (queued messages pending).

Within a session, individual user interactions are executed as \emph{runs}.
A run is a transient execution instance identified by a \texttt{runId}. Run progression can be traced through debug log messages that mark stages such as \texttt{prompt start}, \texttt{agent start}, tool execution, \texttt{agent end}, and \texttt{prompt end}. In parallel, a structured event system emits events on categorized streams: the \texttt{lifecycle} stream tracks run state via phases \texttt{start}, \texttt{end}, and \texttt{error}; the \texttt{tool} stream records tool invocations (\texttt{start}, \texttt{update}, \texttt{result}); and the \texttt{assistant} stream carries model output. Together, these two trace layers enable fine-grained forensic reconstruction of individual runs.

\paragraph{Session Index and Mapping.}
Local session data is stored under \texttt{\textasciitilde/.openclaw/agents/<agentId>/sessions/}.
Within this directory, \texttt{sessions.json} tracks current and past sessions and maps \texttt{sessionKey} values to their corresponding \texttt{sessionId} values.
The \texttt{sessionId} can then be used to locate the session transcript file \texttt{<sessionId>.jsonl}.

\paragraph{Session Transcript Files.}
Session history is persisted in JSONL files located at \noindent \texttt{\textasciitilde/.openclaw/\-agents/\-<agentId>/\-sessions/\-<sessionId>.jsonl}, where each line is a self-contained JSON object with \texttt{id} and \texttt{parentId} fields. The file begins with a \texttt{session} header entry (version 3) containing the session UUID, creation timestamp, and working directory.
Subsequent entries record configuration changes, extension-injected state, and---most importantly---the conversation itself as \texttt{message} entries carrying a \texttt{role} field and typed content blocks. The specific record types are described alongside the components they relate to in the following subsections.

\paragraph{Lifecycle and Recovery.}
OpenClaw employs a soft-delete mechanism: when a session is deleted, the transcript is renamed in place with a \texttt{.deleted.<timestamp>} suffix and removed from the \texttt{sessions.json} index, rather than being expunged.

\subsubsection{Log Data}
\label{subsubsec:logs}
Log data is stored in \texttt{/tmp/openclaw/}. The directory may contain multiple log files, typically one per day, named \texttt{openclaw-YYYY-MM-DD.log}. The retention policy is hard-coded: log files older than 24 hours are deleted. The log cleanup routine runs automatically when the logger is first initialized (early in application startup) or when the logger configuration changes.

Log data is stored in JSON format, with each line representing an independent JSON object that includes details about the subsystem that generated the entry, along with a \texttt{time} field that records when it was emitted.

\subsection{Agent Characterization}
\label{subsec:agent_characterization}
The first step in examining an agentic system is to establish what the agent is, how it reasons, and what it could do. This requires answering several investigative questions: How is the agent configured? What \ac{LLM} drives its reasoning? What capabilities and permissions are available? How has the configuration changed over time?

\subsubsection{Configuration and Identity}
The primary configuration is stored in \texttt{openclaw.json} under the \texttt{agents} section. It defines defaults such as the selected model (e.g., Gemini 2.5 Flash), as well as global settings including the workspace root, which defaults to \texttt{workspace/} within \texttt{\textasciitilde{}/.openclaw/}. During onboarding, OpenClaw creates an initial agent named \texttt{main}; additional agents can be added later and are listed in the \texttt{list} entry. Each agent entry specifies, among other parameters, the model, workspace path, and an agent directory.

The agent's persona and operational boundaries are defined by a set of workspace files initiated by \texttt{AGENTS.md} and \texttt{BOOTSTRAP.md} (the latter is deleted after initialization by the agent itself). Crucial files include: \texttt{IDENTITY.md} (agent name/avatar/``creature type''), \texttt{SOUL.md} (core truths, personality traits, and safety boundaries), \texttt{TOOLS.md} (environment-specific technical details, e.g., SSH hosts and aliases), and \texttt{HEARTBEAT.md} (periodic awareness tasks executed in the main session). These files, together with the system prompt assembled from them, constitute the agent's configured identity and behavioral policy.

Agent-specific provider credentials (e.g., API keys or OAuth tokens) are stored in \texttt{auth-profiles.json} within the respective agent directory. %
Session-level configuration and operational metadata can be reconstructed from \texttt{agents/\{id\}/sessions/sessions.json}, which may record per-session parameters such as the active provider/model as well as usage statistics (e.g., token utilization).

\subsubsection{Capabilities and Permissions}
OpenClaw distinguishes between \emph{tools} and \emph{skills}. Tools represent executable actions (e.g., reading/writing files or invoking \texttt{exec} to run commands), whereas skills provide task-specific guidance and prompt material to teach the agent how to use tools effectively.

Skills are defined by a \texttt{SKILL.md} file and can be stored either globally in \texttt{skills/} or agent-locally under \texttt{workspace/skills/}. In addition, OpenClaw may provide bundled skills (e.g., guidance for querying weather information). Tool implementations reside in the OpenClaw codebase (e.g., \texttt{src/agents/tools/*.ts}).

Evidence of tool and skill availability can be extracted from an agent's \texttt{sessions.json}. For skills, the \texttt{skillsSnapshot} structure may include (i) a list of skills, (ii) \texttt{resolvedSkills} with the resolved on-disk path to each \texttt{SKILL.md} and its \texttt{source} (e.g., \texttt{openclaw-bundled}), and (iii) the generated prompt text that was provided to the model. In practice, we observed that the \texttt{systemPromptReport} prompt-reporting structure may not persist reliably over time. When present, it can additionally summarize exposed tools by name and provide schema/summary length metrics; however, it does not preserve the full tool descriptions as transmitted to the model, nor does it document the concrete implementation behind an exposed tool. The same structure also records injected workspace files, as discussed in \cref{subsubsec:context_reconstruction}.

Scheduled automation capabilities are defined in \texttt{cron/jobs.json}; each job captures the schedule (one-shot \texttt{at}, fixed-interval \texttt{every}, or \texttt{cron} expression with optional timezone) and how it is executed, along with payload and runtime state. These definitions reveal what autonomous actions the agent was \emph{configured} to perform. Run history for these jobs is analyzed in \cref{subsec:action_reconstruction}.

\subsubsection{LLM Core and Reasoning}
\label{subsubsec:llm_reasoning}
A distinguishing feature of agentic AI systems is that the agent's decision-making is mediated by an \ac{LLM} whose reasoning process produces forensically recoverable traces. These artifacts have no direct analog in the forensic analysis of traditional, rule-based software. Assistant messages in the transcript additionally record usage metadata---token counts, cost breakdowns, \texttt{stopReason} (e.g., \texttt{stop} or \texttt{toolUse}), and provider-specific data---enabling investigators to quantify model utilization per interaction.

The \ac{LLM} model identity---including provider, model version, and thinking level---is recorded in \texttt{openclaw.json} and per session in \texttt{sessions.json}. Mid-session model changes are captured as \texttt{model\_change} events in the session transcript.
Reasoning traces appear as \texttt{thinking} content blocks within assistant messages. These chain-of-thought blocks may reveal how the agent interpreted a request, which alternatives it considered, and why it selected a particular tool.

The agent layer here is not the source of these traces---it only records what it receives. OpenClaw persists whatever the model provider returns through its API, and the content, granularity, and even the presence of reasoning traces are determined upstream. We observed that it depends on the configured model whether a full chain-of-thought as plaintext alongside the final answer is included or not. For instance, providers of current flagship models like Claude or Gemini have chosen to not expose verbatim thoughts but rather summaries and \emph{thought signatures}, described as ``encrypted representation of the model's internal reasoning state''.\footnote{See \url{https://ai.google.dev/gemini-api/docs/thinking\#signatures}, accessed on 15.07.2026.} Hence, in these cases, investigators examining an agent installation are not analyzing the reasoning of the agent in any direct sense; they are analyzing the partial view of that reasoning that the provider chose to emit at the time of the interaction. Nevertheless, for the model \emph{Gemini 2.5 Flash}, we found the thought summaries to be insightful and comprehensive.

\subsubsection{Historical Configuration}
Historical configurations can be recovered from the configuration backup files (e.g., \texttt{openclaw.json.bak}, \texttt{openclaw.json.bak.1}, \dots). These snapshots preserve prior configurations alongside metadata such as \texttt{meta.lastTouchedAt} and \texttt{meta.lastTouchedVersion}, enabling approximate point-in-time reconstruction: the configuration ``in effect'' for a given time window can be approximated by selecting the most recent backup whose \texttt{lastTouchedAt} precedes that time. Differencing consecutive snapshots further reveals what changed.

Because the effective system prompt and capability envelope may not be stored persistently, investigators should consider complementary sources such as \textit{log data}. In our analysis, logs contained specific \texttt{google tool schema snapshot} entries that enumerate the tools exposed to the model for a run. While these entries are not explicitly linked to a particular session, they are typically emitted immediately after a run starts and can therefore often be associated with the corresponding session via temporal proximity. As logs are continuous, such snapshots can help identify capability changes over time and detect orphaned tools that were previously available but have since been removed. Analysts should note that snapshots may be recorded in provider-specific schema formats (e.g., Google tool schema), which can vary across model providers.

\subsection{Knowledge Reconstruction}
\label{subsec:knowledge_reconstruction}
Having established the agent's identity, reasoning engine, and capabilities, the investigator next reconstructs what the agent knew. This addresses questions such as: What information had the agent accumulated? What was in its episodic and long-term memory? What context was available to the \ac{LLM} at the time of the events under investigation?

Because an agent's behavior is shaped not only by its configuration and user input but also by its evolving memory, loaded workspace files, and dynamically available tools, the same request can produce different outcomes at different points in time. The goal is therefore to approximate, as closely as the available evidence allows, the knowledge state that was active during the relevant period.

\subsubsection{Episodic and Long-term Memory}
OpenClaw implements memory primarily as plain Markdown files in the agent workspace. \texttt{workspace/memory/YYYY-MM-DD.md} serves as an append-only daily log; updates to this file are triggered by user requests such as ``remember this''. At session start, the agent is instructed (via \texttt{AGENTS.md}) to read the two most recent daily memory files. The curated \texttt{MEMORY.md} is injected into the system prompt by the platform.

OpenClaw can additionally build a vector index over memory files to support semantic retrieval, stored in \texttt{memory/<agent\_id>.sqlite}.

\subsubsection{Contextual Knowledge at Incident Time}
\label{subsubsec:context_reconstruction}
A central challenge is reconstructing what the agent knew \emph{at a specific point in time}. The \texttt{USER.md} file, for instance, contains learned user details such as timezone, active projects, and preferences---information that accumulates and evolves as the agent interacts with the user.

When present, the \texttt{systemPromptReport} in \texttt{sessions.json} (cf.\ \cref{subsec:agent_characterization}) may include an \texttt{injectedWorkspaceFiles} list (with \texttt{name}, \texttt{path}, and injected character counts), which helps approximate which workspace and memory files were injected into the model context. However, this structure may not persist reliably, may be missing entirely, or may only reflect the latest recorded state rather than every individual interaction.

One complementary approach is to reconstruct historical versions of knowledge files from the session transcript itself. When the agent uses file-manipulation tools such as \texttt{read}, \texttt{write}, or \texttt{edit}, the session JSONL records the invoked operation and its payload. This allows investigators to recover prior file contents as last observed or modified by the agent and to build an approximate timeline of context evolution driven by the \ac{LLM}. However, this approach cannot account for \emph{external} modifications to injected files that occur outside the agent's tool interface (e.g., direct user edits in the workspace).

For long-running sessions, investigators should also pay attention to \texttt{compaction} entries in the transcript. These indicate where the context window was summarized and mark locations where the agent's context may have changed.
\subsection{Interaction Tracing}
\label{subsec:interaction_tracing}
The investigator next reconstructs the communication between users and the agent. This requires answering: Through which channels could a user interact with the agent? What and when did the user communicate? What information was visible to the user at any given point?

Together, these findings establish the human side of the interaction and define the boundary between what was requested and what the user could observe.

\subsubsection{Communication Channels}
Channel configuration is centralized in the \texttt{channels} section of \texttt{openclaw.json}. It defines, for each supported channel (e.g., Telegram, WhatsApp, Discord), whether the channel is enabled and which access policies apply.

Because OpenClaw uses channel-specific implementations---for example, the Telegram Bot API for Telegram and the Baileys library for WhatsApp---the resulting artifacts and their storage locations are likewise channel-specific. For Telegram, the bot token is stored directly in \texttt{openclaw.json}, while \texttt{credentials/telegram-allowFrom.json} stores the \texttt{allowFrom} entry used for permission control. For WhatsApp, credentials and auxiliary material (e.g., pre-keys) are stored under \texttt{credentials/whatsapp/default/}.

In addition, \texttt{agents/<id>/sessions/sessions.json} can contain per-session channel metadata, such as the current and last \texttt{channel} as well as the last recipient identifier (e.g., \texttt{telegram:id}). \texttt{sessions.json} may optionally record an origin entry that stores the channel in \texttt{provider} and carries channel-specific information, e.g., sender identifier in the \texttt{from} field or a conversation \texttt{label}, which can be correlated to user messages in a transcript file. Log entries from subsystems such as \texttt{gateway} or channel-specific subsystems (e.g., \texttt{gateway/channels/telegram}) document the connection state and operational health of communication channels over time. Furthermore, comparing configuration backups enables approximate historical reconstruction of channel enablement and policy changes.

\subsubsection{Message Recovery and Media}
Within the session JSONL files, user-to-agent messages are indicated by \texttt{role: user} and include the full text content as well as a timestamped event record.

For some channels, additional attribution is embedded in message headers. For example, Telegram messages include a structured header inside the stored user message content (e.g., \texttt{[Telegram <name> (@<handle>) id:<id> <YYYY-MM-DD> <HH:MM> UTC]}). When present, it can be parsed to extract channel attribution, sender identity (e.g., display name, username, platform-specific user ID), and message timestamp. A platform message identifier appears on a separate line as \texttt{[message\_id: N]}.

When a user sends media attachments, OpenClaw stores inbound files under \texttt{\textasciitilde/.openclaw/media/inbound/} using one of two filename formats: if the channel provides an original filename, it becomes \texttt{\{sanitized\_original\}---\{uuid\}.\{ext\}}; otherwise, only the UUID is used. The extension is determined by MIME type detection. Corresponding session events include a note in the format \texttt{[media attached: PATH (mime/type) | URL]}, allowing media files to be linked back to the conversational timeline.

\subsubsection{Visibility to the User}
The transcript also contains assistant-side events (\texttt{role: assistant}) that may contain multiple content types (e.g., tool-related messages, intermediate reasoning, and user-facing text). According to the implementation, for providers that use reasoning tags (e.g., Google Gemini), only content of type \texttt{text} wrapped in \texttt{<final>} tags is forwarded to the user, while internal thinking blocks (wrapped in \texttt{<think>} tags) and tool execution details are not displayed. For providers with native thinking support, this filtering is not enforced. Accordingly, forensic reconstruction should distinguish between the full internal transcript stored on disk and the subset of content that was actually observable to the user.

\subsection{Action Reconstruction}
\label{subsec:action_reconstruction}
This phase focuses on the actions the agent performed. The investigator constructs a timeline of tool invocations, file operations, network requests, scheduled tasks, and other executed actions. Key questions include: What is the timeline of executed actions? Did the agent execute unprompted or autonomous actions?

The output of this phase is a factual record of what the agent did and when, without yet attempting to explain why.

\subsubsection{Tool Invocations and Effects}
Tool executions are recorded as paired events in the session JSONL files: \texttt{toolCall} content blocks in assistant messages specify the tool name, a unique call identifier (\texttt{id}), and the arguments; subsequent \texttt{role: toolResult} messages reference this identifier as \texttt{toolCallId} and contain the output in \texttt{content}, an \texttt{isError} flag, and a \texttt{timestamp}. For shell commands (\texttt{exec} tool), an optional \texttt{details} object may include \texttt{durationMs}, \texttt{exitCode}, and \texttt{status}.

Log files are an additional source as they contain specific entries for tool start and end that include \texttt{runId} and \texttt{toolCallId}, enabling cross-source correlation. Anti-forensics indicators include tool calls present in logs but absent from the session transcript, which can suggest session-file tampering.

\subsubsection{Autonomous and Scheduled Actions}
Run history for scheduled tasks is appended to \texttt{cron/runs/<jobId>.jsonl} with timestamps and outcomes, enabling investigators to flag activity that lacks an immediately preceding user request. Each job may be executed either in the main session via an injected system event or in an isolated \texttt{cron:<jobId>} run; this distinction determines whether cron-triggered actions appear in the main session transcript or in separate session files.

In addition, the agent may autonomously delegate work to subagents. Delegation can be reconstructed from \texttt{sessions\_spawn} tool calls (task prompt, child session ID, cleanup policy, token use, and result summary) and from \texttt{sessions.json}, which may record the parent in \texttt{spawnedBy} and reference the child under keys such as \texttt{agent:main:subagent:<uuid>}. If the cleanup policy deletes the entry from \texttt{sessions.json}, the child session transcript often remains on disk, allowing investigators to link parent and child via the recorded session identifiers and to attribute seemingly user-initiated events (\texttt{role: user}) in the subagent transcript to agent-driven spawning.

OpenClaw additionally maintains a lightweight registry of spawned subagent runs in \texttt{subagents/runs.json}. It records high-level metadata such as which subagent was spawned for which task and when it started and ended, which can be useful when session records are incomplete. Entries are created when a subagent is spawned and are updated as lifecycle events occur; depending on the configured cleanup policy, they may be removed immediately after completion or retained only temporarily before being swept once their \texttt{archiveAtMs} deadline passes (defaulting to 60~minutes after creation), so their presence (or absence) can itself be an indicator of how the system was configured and operated.

\subsection{Attribution Analysis}
\label{subsec:attribution}
The preceding phases recover evidence about what the agent was, how it reasoned, what it knew, how it communicated, and what it did. This final phase synthesizes evidence \emph{across} all areas to address questions of causality and responsibility: Can specific actions be linked to specific user requests? What degree of autonomy was exhibited? What are the limits of the reconstruction?

Unlike the evidence-recovery phases, attribution is an \emph{interpretive} activity: it does not recover new artifacts but reasons about the relationships between artifacts already identified.

\subsubsection{Linking Actions to Origins}
The investigator traces each significant action backward through the session transcript to determine its origin. Did the reasoning trace cite a user message? A memory entry? A cron trigger? Its own prior reasoning? The chain from user request through agent reasoning to tool invocation can often be reconstructed from the sequence of \texttt{role: user}, \texttt{role: assistant} (with \texttt{thinking} blocks), and \texttt{toolCall}/\texttt{toolResult} entries within a session.

\subsubsection{Autonomy Assessment}
Indicators of autonomy are distributed across multiple artifact sources. Cron run history quantifies scheduled actions that occurred without a preceding user message. Subagent delegation reveals autonomous task decomposition. Each observed action can be classified on a spectrum: directly instructed by the user, interpretively derived from an ambiguous instruction, autonomously initiated by the agent, or indeterminate. This classification draws on evidence from all preceding phases---the agent's configuration and reasoning capabilities, its knowledge context, the communication record, and the action timeline.

\subsubsection{Reconstruction Boundaries}
Several factors limit the completeness of any forensic reconstruction of an agentic AI system. First, the \ac{LLM}'s reasoning is only partially observable through thinking traces. Second, the actual context window assembled at inference time is not stored persistently and memory as well as configuration may have changed between the incident and the examination. Lastly, as we will discuss in \cref{sec:discussion}, there is an inherent nondeterminism of \ac{LLM}-based decision-making, so even a complete reconstruction of inputs cannot guarantee a unique explanation for the agent's behavior.

\section{A Forensic Taxonomy for Agentic AI} %
\label{sec:taxonomy}
AI agents are notably different from regular rule-based software systems, so there is a need for the research community to provide guidance for analyzing agent-specific components such as context, configuration, execution, \ac{LLM} interaction, and autonomous scheduling. To address this gap, we structure the findings of our OpenClaw analysis into an \emph{Agent Artifact Taxonomy}---a functional classification of five artifact planes, each corresponding to a distinct aspect of the agent's architecture. The taxonomy was developed through a combination of literature review on agentic system architectures, code analysis as well as differential analysis, and validated against the abstract agent interaction loop described in \cref{sec:bg_related_work:single_agent}.

\subsection{Five Artifact Planes}
\label{subsec:five_planes}
\begin{figure*}[htbp]
  \centering
  \resizebox{0.98\textwidth}{!}{
    \includegraphics{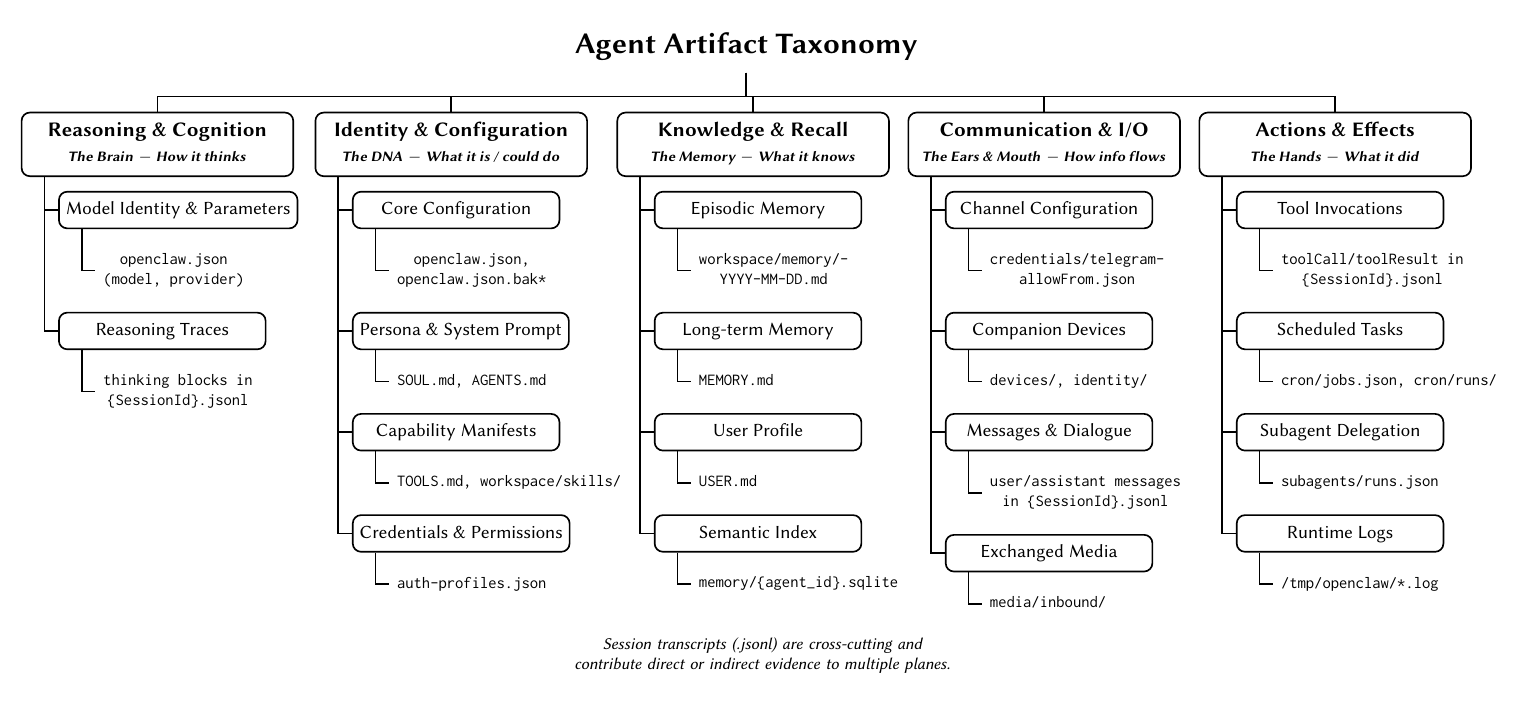}
  }
  \caption{Agent Artifact Taxonomy for forensic analysis of agentic AI systems. The taxonomy organizes agent-related evidence into five planes, each corresponding to a distinct aspect of the agent's architecture. Session transcripts are cross-cutting and contribute direct or indirect evidence to multiple planes. Representative OpenClaw artifacts are listed as examples.}
  \Description{Tree diagram: the root node, Agent Artifact Taxonomy, branches into five planes. Reasoning and Cognition (the brain) covers model identity and parameters as well as reasoning traces, exemplified by thinking blocks in session transcripts. Identity and Configuration (the DNA) covers core configuration, persona and system prompt, capability manifests, and credentials and permissions, exemplified by openclaw.json, SOUL.md, and auth-profiles.json. Knowledge and Recall (the memory) covers episodic memory, long-term memory, user profile, and semantic index, exemplified by daily memory files, MEMORY.md, USER.md, and the memory SQLite database. Communication and I/O (the ears and mouth) covers channel configuration, companion devices, messages and dialogue, and exchanged media. Actions and Effects (the hands) covers tool invocations, scheduled tasks, subagent delegation, and runtime logs, exemplified by toolCall records, cron jobs, subagent runs, and log files. A note marks session transcripts as cross-cutting evidence contributing to multiple planes.}
  \label{fig:taxonomy}
\end{figure*}
Just as one might examine attributes indicating a suspect's intent with the cognitive processes, inherent characteristics, knowledge, communications, and actions, the five planes capture analogous aspects of an AI agent. \Cref{fig:taxonomy} provides an overview with representative artifacts; we briefly introduce each plane below.

\begin{enumerate}[leftmargin=*, itemsep=3pt, topsep=3pt, label=\textbf{\arabic*.}]
  \item \textbf{Reasoning \& Cognition (``The Brain'')}: \emph{how the agent thinks.} Captures the \ac{LLM}'s reasoning traces: a category of forensic evidence with no analog in traditional software and the closest approximation to reconstructing agent \textit{intent}.

  \item \textbf{Identity \& Configuration (``The DNA'')}: \emph{what the agent is and what it could do.} Encompasses the predominantly \emph{static} artifacts that define the agent's configured identity, capabilities, and operational boundaries.

  \item \textbf{Knowledge \& Recall (``The Memory'')}: \emph{what the agent knows.} Covers accumulated knowledge such as episodic memory, user profiles, and semantic indices, distinguished from DNA by being \emph{accumulated at runtime}, often by the agent itself.

  \item \textbf{Communication \& I/O (``The Ears \& Mouth'')}: \emph{how information flows to and from the agent.} Establishes who interacted with the agent, through which channels, and what each party could observe.

  \item \textbf{Actions \& Effects (``The Hands'')}: \emph{what the agent did.} Records concrete actions and their observable effects, providing the factual basis for determining what changes the agent caused.
\end{enumerate}

The planes represent analytical perspectives rather than isolated subsystems, and individual artifacts may participate in multiple planes depending on investigative context. In particular, session transcripts are inherently multi-planar: a single JSONL file directly records reasoning traces (Brain), user and agent messages (Ears~\&~Mouth), and tool invocations (Hands), while additionally containing indirect evidence for the DNA and Memory planes.

\subsection{Preliminary Cross-System Assessment}
\label{sec:preliminary_validation}
The analytical phases applied in \cref{sec:analysis} map closely to these planes, reflecting the modular architecture of agentic AI systems. While the specific artifacts differ across systems, the taxonomy is designed to generalize: any agent that combines an \ac{LLM} core with persistent configuration, memory, communication interfaces, and tool execution will produce artifacts classifiable into these five planes.
For a preliminary validation, we mapped the taxonomy's five planes against artifacts reported by~\citet{WalkerGAHB24} for the agentic framework AutoGen and by~\citet{DragonasLN24} for the non-agentic ChatGPT mobile application. All AI-specific artifacts documented in both studies fit within the existing planes without requiring additions, supporting the taxonomy's structural completeness detailed in \cref{tab:cross_validation}.
The mapping reveals a discriminative pattern: the Reasoning \& Cognition and Actions \& Effects planes are both effectively empty for the non-agentic ChatGPT, while the Identity, Knowledge, and Communication planes are populated---mirroring the architectural distinction between systems that merely \emph{use} an \ac{LLM} and those that \emph{reason and act} via one. AutoGen, despite being a genuine multi-agent framework, shows a similar gap. Grounded in traditional disk, memory, and network forensics, \citet{WalkerGAHB24} did not consider reasoning traces or structured action records as evidence categories; combined with the reported deprecation of runtime logging since AutoGen~v0.2, this left both the Reasoning and Actions planes without recoverable artifacts. This demonstrates how agent-specific evidence classes can go unexamined without a prescriptive taxonomy that defines which artifact categories investigators must consider---ensuring that fundamentally new classes such as reasoning traces are not overlooked.

\begin{table*}[htbp]
  \centering
  \caption{Literature-based cross-system mapping of the Agent Artifact Taxonomy. ChatGPT mobile is a cloud-native \ac{LLM} chat interface without agent capabilities per \cref{sec:bg_related_work:single_agent}; it is included as an architectural contrast to illustrate the taxonomy's discriminative scope.}
  \label{tab:cross_validation}
  \scriptsize
  \begin{tabularx}{\textwidth}{@{} l X X X @{}}
    \toprule
    \textbf{Taxonomy Plane} & \textbf{OpenClaw (this work)} & \textbf{AutoGen (\citet{WalkerGAHB24})} & \textbf{ChatGPT Mobile (\citet{DragonasLN24})} \\
    \midrule
    Reasoning \& Cognition
      & \texttt{thinking} blocks and \texttt{model\_change} events in session JSONL; reasoning traces with no analog in traditional software
      & No reasoning traces recovered or investigated; transient LLM response fragments in volatile memory but recovery inconsistent across tasks; no chain-of-thought, planning step, or inference parameter artifacts
      & No local reasoning traces (all inference possibly server-side); GPT model version recorded in conversation metadata but contains no chain-of-thought or inference parameters \\
    \addlinespace
    Identity \& Configuration
      & \texttt{openclaw.json}, persona files (\texttt{SOUL.md}, \texttt{IDENTITY.md}), provider credentials in \texttt{auth-profiles.json}
      & Agent definitions, system prompt, model configuration (\texttt{llm\_config}), and code execution settings (\texttt{work\_dir}) in Python source scripts; \texttt{pyautogen} installation footprint in \texttt{site-packages}; API credentials not recovered
      & App preferences (BPLIST on iOS, Protobuf on Android), account/user/device/workspace identifiers, subscription plan, custom instructions, authentication tokens \\
    \addlinespace
    Knowledge \& Recall
      & \texttt{memory.sqlite} with vector embeddings, curated \texttt{MEMORY.md}, daily memory logs, \texttt{USER.md}
      & Conversational context session-scoped and volatile only (not persisted to disk); logging deprecated since v0.2; no persistent memory, no user profiles, no semantic indices
      & Conversation SQLite database (Android) and JSON files (iOS); cloud-synced chat history; no persistent agent memory; no local semantic index \\
    \addlinespace
    Communication \& I/O
      & Multi-channel transcripts (Telegram, WhatsApp), \texttt{media/inbound/} attachments, channel config in \texttt{openclaw.json}
      & TLS-encrypted PCAP traces of API connections to OpenAI endpoints; inter-agent dialogue (UserProxyAgent $\leftrightarrow$ AssistantAgent) not persistently logged; message content encrypted under TLS~1.3 (decryption not attempted)
      & Chat messages in local database and JSON; voice recordings (\texttt{.m4a}); text-to-speech narrations; uploaded images and files; cloud export archive \\
    \addlinespace
    Actions \& Effects
      & \texttt{toolCall}/\texttt{toolResult} pairs in session JSONL, \texttt{cron/jobs.json} and \texttt{cron/runs/} logs, \texttt{subagents/runs.json}
      & No structured tool invocation logs or agent delegation records recovered; runtime logging deprecated since v0.2
      & Minimal: DALL-E image generation among tested actions but no dedicated artifact beyond conversation record; no local tool execution, no scheduled tasks, no autonomous actions \\
    \bottomrule
  \end{tabularx}
\end{table*}

\section{Discussion}
\label{sec:discussion}

\subsection{Practical Implications of Our Analysis}
AI agents expand the attack and impact surface by mediating high-privilege actions (e.g., file system access, process execution, and network requests) through tool abstractions exposed via asynchronous chat interfaces. Furthermore, the intended autonomy---while beneficial for usability---can increase both the likelihood and severity of unintended or adversarial actions, including data exfiltration, credential misuse, and unsafe system modifications, through adversarial content designed to interfere with visiting agents~\citep{FranklinTJLO26} or compromised agent instances.
This interaction model weakens real-time human oversight: tool calls may execute without immediate user confirmation, and their side effects remain unnoticed. The autonomous nature introduces a covert communication risk that differs from supervised interactive sessions. This shifts the trust model from visible session oversight to post-hoc audit and enables ``silent informant'' behavior: an assistant completes a benign task while covertly transmitting underlying data.

In practice, the artifact catalog and taxonomy derived from our controlled analysis directly support field examinations. The taxonomy's five planes provide a structured checklist of evidence categories, while the specific file paths, data formats, and content semantics documented in \cref{sec:analysis} serve as a forensic reference that guides examiners to relevant traces without prior knowledge of the application.

\subsection{Limitations of This Study}
\label{sec:limitations}
During our analysis period, the OpenClaw codebase evolved rapidly. While this is a general phenomenon in digital forensics~\citep{SpichigerA25}, the fast-paced development style rooted in ``vibe coding practices''~\citep{MeskeHWLB25} and the increasing autonomy of machines creating software aggravates the situation and makes OpenClaw a fast moving target for empirical study.
Accordingly, we restricted our examination to persistent on-disk artifacts and excluded main memory and network traffic. We also did not assess the durability of this evidence over extended periods. Moreover, we focused on core functionality using a set of representative actions and did not consider companion devices. Our experiments were conducted on a single operating system (Debian Linux). While the application-layer artifacts are platform-independent (cf.\ \cref{sec:methodology}), OS-level traces such as file system metadata semantics, journal entries, or package-manager records will differ across platforms and warrant dedicated cross-platform validation.
The \ac{LLM}'s inherent nondeterminism complicated reproducible experiments: despite fixing all variables under our control, we could not make OpenClaw behave completely identically across runs. This highlights that agentic systems complicate reproducible analysis compared to traditional, non-agentic software.
The proposed taxonomy is inductively derived from our analysis of OpenClaw. A preliminary literature-based cross-mapping against published artifact inventories from two additional studies supports the taxonomy's structural completeness (\cref{tab:cross_validation}), but hands-on forensic analysis of additional agentic systems is needed to fully validate its generalizability.

\subsection{The Need for an Agentic AI Investigation Framework}
\label{sec:need_for_agentic_ai}
\paragraph{Nondeterminism.}
Our analysis revealed that repeated execution of identical actions with fixed conditions does not yield identical traces.
For example, issuing a reminder request may either create a cron job (in \texttt{\textasciitilde/.openclaw/cron/}) or instead modify \texttt{HEARTBEAT.md}, depending on the context envelope provided to the \ac{LLM}. %
We distinguish three sources of nondeterminism: (1) sampling stochasticity: the \ac{LLM} produces different reasoning chains and tool selections for the same prompt; (2) context assembly variability: the dynamically constructed context window incorporating evolving memory files and workspace content differs subtly between runs; and (3) timing-dependent external state: responses from third-party services are inherently non-reproducible. Crucially, not all artifacts are equally affected. Structural artifacts, such as configuration files, credential stores, directory layouts, and file creation patterns, remained deterministic across our runs, while LLM-generated content, i.e., reasoning traces, tool-choice sequences, and agent-authored memory entries, exhibited variation.

\paragraph{Abstraction.}
These observations suggest two qualitatively different classes of traces: (i) conventional, deterministic artifacts produced by rule-based software components, and (ii) agent-generated artifacts whose creation depends on \ac{LLM}-mediated planning and tool choice.
The latter class has important implications for reconstruction and attribution, because the same high-level intent can be realized through different execution paths and, consequently, different local traces, which can be conceptualized as an additional abstraction layer~\citep{Carrier03} in trace creation.

\paragraph{Challenges in Context Reconstruction.}
Our experiments showed that it is highly relevant to determine which context was available to the \ac{LLM} at what point in time. Since the workspace environment, memory, and capabilities naturally evolve and change over time, this is difficult to determine. The correlation of modified timestamps of the context-providing files in the workspace or memory directory might give an indication but cannot provide the full picture either since many relevant parameters are difficult to determine retrospectively.
A practical way to reduce this interpretive uncertainty is to recover as many parameters as possible that constrain the execution environment (e.g., available tools and permissions). In practice, however, some of these parameters are difficult to determine retrospectively; for instance, capability availability is not logged explicitly.

Closely related are difficulties in reconstructing the agent's intent.
While full access to the underlying reasoning process has been shown to provide supporting details for understanding, interpreting, and validating outputs in digital forensics tasks~\citep{MicheletSWB26}, the trend over the course of 2025 toward encrypting the internal thought process and exposing only model-generated summaries\footnote{See \url{https://docs.claude.com/en/docs/build-with-claude/extended-thinking}, \url{https://platform.openai.com/docs/guides/reasoning}, \url{https://cookbook.openai.com/examples/responses_api/reasoning_items}, \url{https://ai.google.dev/gemini-api/docs/thinking\#signatures}, all accessed on 15.07.2026.} limits what is available in practice.
Moreover, the visible summaries that remain cannot necessarily be treated as fully faithful records of the underlying reasoning, as providers' own evaluations and independent analyses have shown that reasoning models frequently fail to verbalize the cues that actually shaped their answers~\citep{chen2025reasoning,arcuschin2025chainofthought}.

Generally, we see the need for clear guidance on the task of context reconstruction; our taxonomy proposal in \cref{sec:taxonomy} may provide a viable foundation, but it still requires further development.

\section{Future Work}
\label{sec:future_work}
Future work includes extending the analysis to additional AI assistants and deployment settings with different communication channels and OSes.
We further plan to broaden the set of covered trace classes (e.g., main memory and network-related artifacts) and to apply and potentially refine the taxonomy and workflow across platforms, diverse configurations and usage scenarios.
Reconstructing model reasoning from encrypted blobs remains methodologically open. Future work should explore model-assisted decoding of thought signatures and inference from residual traces across the Configuration, Knowledge, Communication, and Action planes.
More fundamentally, we identify risks to digital forensic capabilities arising from (i) the accelerated pace of software production (e.g., via ``vibe coding'') and (ii) the absence of robust forensic readiness measures, which may hinder the reliable investigation of actions performed by AI agents with full system privileges.

\section{Conclusion}
\label{sec:conclusion}
Agentic AI systems are becoming a durable part of the software ecosystem. As they increasingly act on behalf of users
they will reshape digital processes and become a consequential object of digital investigations. At the same time, their delegated and context-dependent execution challenges long-standing forensic assumptions about reproducibility, provenance, and the completeness of local traces.

In this work, we present, to our knowledge, the first forensic analysis of a widely used personal AI assistant, taking \emph{OpenClaw} as our case study. We combined an architectural perspective with static code analysis and differential forensic analysis to identify recoverable on-disk traces and to relate these traces to agent-relevant investigative questions, for which we provide an analysis tool as open-source software. We further generalized our observations by proposing an artifact taxonomy with five planes: reasoning and cognition, identity and configuration, knowledge and recall, communication and I/O, and actions and effects---to structure investigations of agentic assistants. Our findings also highlight why agentic systems cannot be approached like traditional software. First, agentic execution introduces nondeterminism: the same user-level request may be realized through different tool choices and different state transitions, producing different artifacts. Second, \acp{LLM} add an additional layer of abstraction between intent, actions, and observable traces, and third, the relevant context that shaped a decision is difficult to reconstruct retrospectively.
Overall, agentic AI shifts the conditions of analysis, compelling a reassessment of long-standing assumptions in a landscape shaped by delegated agency, autonomy, and model-mediated execution, underscoring the need for sustained, systematic forensic research of agentic systems.

\begin{acks}
  Jan Gruber was funded by the Deutsche Forschungsgemeinschaft (DFG, German
  Research Foundation) -- 570978749.
\end{acks}

\section*{Use of AI writing assistance}
The authors used Claude Opus 4.8 and Gemini 3.1 Pro as writing aids to identify typographical and grammatical issues and to improve the clarity and style of selected passages. All suggestions were critically assessed and, where appropriate, adapted by the authors.

\bibliographystyle{ACM-Reference-Format}
\bibliography{references}


\begin{thebibliography}{29}


\ifx \showCODEN    \undefined \def \showCODEN     #1{\unskip}     \fi
\ifx \showDOI      \undefined \def \showDOI       #1{#1}\fi
\ifx \showISBNx    \undefined \def \showISBNx     #1{\unskip}     \fi
\ifx \showISBNxiii \undefined \def \showISBNxiii  #1{\unskip}     \fi
\ifx \showISSN     \undefined \def \showISSN      #1{\unskip}     \fi
\ifx \showLCCN     \undefined \def \showLCCN      #1{\unskip}     \fi
\ifx \shownote     \undefined \def \shownote      #1{#1}          \fi
\ifx \showarticletitle \undefined \def \showarticletitle #1{#1}   \fi
\ifx \showURL      \undefined \def \showURL       {\relax}        \fi
\providecommand\bibfield[2]{#2}
\providecommand\bibinfo[2]{#2}
\providecommand\natexlab[1]{#1}
\providecommand\showeprint[2][]{arXiv:#2}

\bibitem[Ali et~al\mbox{.}(2026)]%
        {AliDC26}
\bibfield{author}{\bibinfo{person}{Mohamad~Abou Ali}, \bibinfo{person}{Fadi
  Dornaika}, {and} \bibinfo{person}{Jinan Charafeddine}.}
  \bibinfo{year}{2026}\natexlab{}.
\newblock \showarticletitle{Agentic {AI:} a comprehensive survey of
  architectures, applications, and future directions}.
\newblock \bibinfo{journal}{\emph{Artif. Intell. Rev.}} \bibinfo{volume}{59},
  \bibinfo{number}{1} (\bibinfo{year}{2026}), \bibinfo{pages}{11}.
\newblock
\urldef\tempurl%
\url{https://doi.org/10.1007/S10462-025-11422-4}
\showDOI{\tempurl}


\bibitem[Arcuschin et~al\mbox{.}(2025)]%
        {arcuschin2025chainofthought}
\bibfield{author}{\bibinfo{person}{Iv{\'a}n Arcuschin}, \bibinfo{person}{Jett
  Janiak}, \bibinfo{person}{Robert Krzyzanowski}, \bibinfo{person}{Senthooran
  Rajamanoharan}, \bibinfo{person}{Neel Nanda}, {and} \bibinfo{person}{Arthur
  Conmy}.} \bibinfo{year}{2025}\natexlab{}.
\newblock \showarticletitle{Chain-of-Thought Reasoning In The Wild Is Not
  Always Faithful}.
\newblock \bibinfo{journal}{\emph{arXiv preprint arXiv:2503.08679}}
  (\bibinfo{date}{March} \bibinfo{year}{2025}).
\newblock
\urldef\tempurl%
\url{https://doi.org/10.48550/arXiv.2503.08679}
\showDOI{\tempurl}
\newblock
\shownote{v4, June 2025}.


\bibitem[Baggili and Behzadan(2020)]%
        {BehzadanB20}
\bibfield{author}{\bibinfo{person}{Ibrahim~M. Baggili} {and}
  \bibinfo{person}{Vahid Behzadan}.} \bibinfo{year}{2020}\natexlab{}.
\newblock \showarticletitle{Founding The Domain of {AI} Forensics}. In
  \bibinfo{booktitle}{\emph{Proceedings of the Workshop on Artificial
  Intelligence Safety, co-located with 34th {AAAI} Conference on Artificial
  Intelligence, SafeAI@AAAI 2020, New York City, NY, USA, February 7, 2020}}
  \emph{(\bibinfo{series}{{CEUR} Workshop Proceedings},
  Vol.~\bibinfo{volume}{2560})},
  \bibfield{editor}{\bibinfo{person}{Hu{\'{a}}scar Espinoza},
  \bibinfo{person}{Jos{\'{e}} Hern{\'{a}}ndez{-}Orallo},
  \bibinfo{person}{Xin~Cynthia Chen}, \bibinfo{person}{Se{\'{a}}n~S.
  {\'{O}}h{\'{E}}igeartaigh}, \bibinfo{person}{Xiaowei Huang},
  \bibinfo{person}{Mauricio Castillo{-}Effen}, \bibinfo{person}{Richard
  Mallah}, {and} \bibinfo{person}{John~A. McDermid}} (Eds.).
  \bibinfo{publisher}{CEUR-WS.org}, \bibinfo{pages}{31--35}.
\newblock
\urldef\tempurl%
\url{https://ceur-ws.org/Vol-2560/paper53.pdf}
\showURL{%
\tempurl}


\bibitem[Carrier(2003)]%
        {Carrier03}
\bibfield{author}{\bibinfo{person}{Brian~D. Carrier}.}
  \bibinfo{year}{2003}\natexlab{}.
\newblock \showarticletitle{Defining Digital Forensic Examination and Analysis
  Tools Using Abstraction Layers}.
\newblock \bibinfo{journal}{\emph{Int. J. Digit. EVid.}} \bibinfo{volume}{1},
  \bibinfo{number}{4} (\bibinfo{year}{2003}).
\newblock
\urldef\tempurl%
\url{http://www.utica.edu/academic/institutes/ecii/publications/articles/A04C3F91-AFBB-FC13-4A2E0F13203BA980.pdf}
\showURL{%
\tempurl}


\bibitem[Chen et~al\mbox{.}(2025)]%
        {chen2025reasoning}
\bibfield{author}{\bibinfo{person}{Yanda Chen}, \bibinfo{person}{Joe Benton},
  \bibinfo{person}{Ansh Radhakrishnan}, \bibinfo{person}{Jonathan Uesato},
  \bibinfo{person}{Carson Denison}, \bibinfo{person}{John Schulman},
  \bibinfo{person}{Arushi Somani}, \bibinfo{person}{Peter Hase},
  \bibinfo{person}{Misha Wagner}, \bibinfo{person}{Fabien Roger},
  \bibinfo{person}{Vlad Mikulik}, \bibinfo{person}{Samuel~R. Bowman},
  \bibinfo{person}{Jan Leike}, \bibinfo{person}{Jared Kaplan}, {and}
  \bibinfo{person}{Ethan Perez}.} \bibinfo{year}{2025}\natexlab{}.
\newblock \showarticletitle{Reasoning Models Don't Always Say What They Think}.
\newblock \bibinfo{journal}{\emph{arXiv preprint arXiv:2505.05410}}
  (\bibinfo{date}{May} \bibinfo{year}{2025}).
\newblock
\urldef\tempurl%
\url{https://doi.org/10.48550/arXiv.2505.05410}
\showDOI{\tempurl}
\newblock
\shownote{Anthropic Alignment Science Team. Companion blog post: ``Reasoning
  models don't always say what they think,'' Anthropic, 3 April 2025,
  https://www.anthropic.com/research/reasoning-models-dont-say-think}.


\bibitem[Chernyshev et~al\mbox{.}(2024)]%
        {ChernyshevBD24}
\bibfield{author}{\bibinfo{person}{Maxim Chernyshev},
  \bibinfo{person}{Zubair~A. Baig}, {and} \bibinfo{person}{Robin Ram~Mohan
  Doss}.} \bibinfo{year}{2024}\natexlab{}.
\newblock \showarticletitle{Towards Large Language Model {(LLM)} Forensics
  Using LLM-based Invocation Log Analysis}. In
  \bibinfo{booktitle}{\emph{Proceedings of the 1st {ACM} Workshop on Large {AI}
  Systems and Models with Privacy and Safety Analysis, {LAMPS} 2024, Salt Lake
  City, UT, USA, October 14-18, 2024}},
  \bibfield{editor}{\bibinfo{person}{Bo~Li}, \bibinfo{person}{Wenyuan Xu},
  \bibinfo{person}{Jieshan Chen}, \bibinfo{person}{Yang Zhang},
  \bibinfo{person}{Jason Xue}, \bibinfo{person}{Shuo Wang},
  \bibinfo{person}{Guangdong Bai}, {and} \bibinfo{person}{Xingliang Yuan}}
  (Eds.). \bibinfo{publisher}{{ACM}}, \bibinfo{pages}{89--96}.
\newblock
\urldef\tempurl%
\url{https://doi.org/10.1145/3689217.3690616}
\showDOI{\tempurl}


\bibitem[Chung et~al\mbox{.}(2017)]%
        {ChungPL17}
\bibfield{author}{\bibinfo{person}{Hyunji Chung}, \bibinfo{person}{Jungheum
  Park}, {and} \bibinfo{person}{Sangjin Lee}.} \bibinfo{year}{2017}\natexlab{}.
\newblock \showarticletitle{Digital forensic approaches for Amazon Alexa
  ecosystem}.
\newblock \bibinfo{journal}{\emph{Digit. Investig.}}  \bibinfo{volume}{22
  Supplement} (\bibinfo{year}{2017}), \bibinfo{pages}{S15--S25}.
\newblock
\urldef\tempurl%
\url{https://doi.org/10.1016/J.DIIN.2017.06.010}
\showDOI{\tempurl}


\bibitem[Crasselt and Pugliese(2024)]%
        {CrasseltG24}
\bibfield{author}{\bibinfo{person}{Jona Crasselt} {and} \bibinfo{person}{Gaston
  Pugliese}.} \bibinfo{year}{2024}\natexlab{}.
\newblock \showarticletitle{{Started Off Local, Now We're in the Cloud:
  Forensic Examination of the Amazon Echo Show 15 Smart Display}}. In
  \bibinfo{booktitle}{\emph{Proceedings of the Digital Forensics Research
  Conference Europe (DFRWS USA)}}, \bibfield{editor}{\bibinfo{person}{Parag
  Rughani} {and} \bibinfo{person}{Frank Breitinger}} (Eds.).
  \bibinfo{publisher}{dfrws.org}, \bibinfo{address}{Baton Rouge, Louisiana},
  \bibinfo{pages}{1--11}.
\newblock
\urldef\tempurl%
\url{https://doi.org/10.48550/ARXIV.2408.15768}
\showDOI{\tempurl}


\bibitem[Dragonas et~al\mbox{.}(2024)]%
        {DragonasLN24}
\bibfield{author}{\bibinfo{person}{Evangelos Dragonas}, \bibinfo{person}{Costas
  Lambrinoudakis}, {and} \bibinfo{person}{Panagiotis Nakoutis}.}
  \bibinfo{year}{2024}\natexlab{}.
\newblock \showarticletitle{Forensic analysis of OpenAI's ChatGPT mobile
  application}.
\newblock \bibinfo{journal}{\emph{Forensic Sci. Int. Digit. Investig.}}
  \bibinfo{volume}{50} (\bibinfo{year}{2024}), \bibinfo{pages}{301801}.
\newblock
\urldef\tempurl%
\url{https://doi.org/10.1016/J.FSIDI.2024.301801}
\showDOI{\tempurl}


\bibitem[F{\"{a}}hndrich et~al\mbox{.}(2023)]%
        {FahndrichHPRBL23}
\bibfield{author}{\bibinfo{person}{Johannes F{\"{a}}hndrich},
  \bibinfo{person}{Wilfried Honekamp}, \bibinfo{person}{Roman Povalej},
  \bibinfo{person}{Heiko Rittelmeier}, \bibinfo{person}{Silvio Berner}, {and}
  \bibinfo{person}{Dirk Labudde}.} \bibinfo{year}{2023}\natexlab{}.
\newblock \showarticletitle{Digital forensics and strong {AI:} {A} structured
  literature review}.
\newblock \bibinfo{journal}{\emph{Forensic Sci. Int. Digit. Investig.}}
  \bibinfo{volume}{46} (\bibinfo{year}{2023}), \bibinfo{pages}{301617}.
\newblock
\urldef\tempurl%
\url{https://doi.org/10.1016/J.FSIDI.2023.301617}
\showDOI{\tempurl}


\bibitem[Franklin et~al\mbox{.}(2026)]%
        {FranklinTJLO26}
\bibfield{author}{\bibinfo{person}{Matija Franklin}, \bibinfo{person}{Nenad
  Tomašev}, \bibinfo{person}{Julian Jacobs}, \bibinfo{person}{Joel~Z. Leibo},
  {and} \bibinfo{person}{Simon Osindero}.} \bibinfo{year}{2026}\natexlab{}.
\newblock \showarticletitle{AI Agent Traps}.
\newblock \bibinfo{journal}{\emph{SSRN Electronic Journal}}
  (\bibinfo{year}{2026}).
\newblock
\urldef\tempurl%
\url{https://doi.org/10.2139/ssrn.6372438}
\showDOI{\tempurl}


\bibitem[Garfinkel(2009)]%
        {Garfinkel09}
\bibfield{author}{\bibinfo{person}{Simson~L. Garfinkel}.}
  \bibinfo{year}{2009}\natexlab{}.
\newblock \showarticletitle{Automating Disk Forensic Processing with SleuthKit,
  {XML} and Python}. In \bibinfo{booktitle}{\emph{Fourth International {IEEE}
  Workshop on Systematic Approaches to Digital Forensic Engineering, {SADFE}
  2009, Berkeley, California, USA, May 21, 2009}}. \bibinfo{publisher}{{IEEE}
  Computer Society}, \bibinfo{pages}{73--84}.
\newblock
\urldef\tempurl%
\url{https://doi.org/10.1109/SADFE.2009.12}
\showDOI{\tempurl}


\bibitem[Garfinkel(2012)]%
        {Garfinkel12}
\bibfield{author}{\bibinfo{person}{Simson~L. Garfinkel}.}
  \bibinfo{year}{2012}\natexlab{}.
\newblock \showarticletitle{Digital forensics {XML} and the {DFXML} toolset}.
\newblock \bibinfo{journal}{\emph{Digit. Investig.}} \bibinfo{volume}{8},
  \bibinfo{number}{3-4} (\bibinfo{year}{2012}), \bibinfo{pages}{161--174}.
\newblock
\urldef\tempurl%
\url{https://doi.org/10.1016/J.DIIN.2011.11.002}
\showDOI{\tempurl}


\bibitem[Garfinkel et~al\mbox{.}(2012)]%
        {GarfinkelNY12}
\bibfield{author}{\bibinfo{person}{Simson~L. Garfinkel},
  \bibinfo{person}{Alex~J. Nelson}, {and} \bibinfo{person}{Joel Young}.}
  \bibinfo{year}{2012}\natexlab{}.
\newblock \showarticletitle{A general strategy for differential forensic
  analysis}.
\newblock \bibinfo{journal}{\emph{Digit. Investig.}} \bibinfo{volume}{9},
  \bibinfo{number}{Supplement} (\bibinfo{year}{2012}),
  \bibinfo{pages}{S50--S59}.
\newblock
\urldef\tempurl%
\url{https://doi.org/10.1016/J.DIIN.2012.05.003}
\showDOI{\tempurl}


\bibitem[Huang and Huang(2025)]%
        {HuangH25b}
\bibfield{author}{\bibinfo{person}{Ken Huang} {and} \bibinfo{person}{Jerry
  Huang}.} \bibinfo{year}{2025}\natexlab{}.
\newblock \bibinfo{booktitle}{\emph{AI Agent Tools and Frameworks}}.
\newblock \bibinfo{publisher}{Springer Nature Switzerland},
  \bibinfo{pages}{23–50}.
\newblock
\showISBNx{9783031900266}
\showISSN{2196-8713}
\urldef\tempurl%
\url{https://doi.org/10.1007/978-3-031-90026-6_2}
\showDOI{\tempurl}


\bibitem[Huang and Hughes(2025)]%
        {HuangH25}
\bibfield{author}{\bibinfo{person}{Ken Huang} {and} \bibinfo{person}{Chris
  Hughes}.} \bibinfo{year}{2025}\natexlab{}.
\newblock \bibinfo{booktitle}{\emph{Introduction to Agentic AI: Foundations,
  Drivers, and Risks}}.
\newblock \bibinfo{publisher}{Springer Nature Switzerland},
  \bibinfo{pages}{3–16}.
\newblock
\showISBNx{9783032021304}
\showISSN{3091-2741}
\urldef\tempurl%
\url{https://doi.org/10.1007/978-3-032-02130-4_1}
\showDOI{\tempurl}


\bibitem[K{\"{a}}lber et~al\mbox{.}(2013)]%
        {KalberDF13}
\bibfield{author}{\bibinfo{person}{Sven K{\"{a}}lber}, \bibinfo{person}{Andreas
  Dewald}, {and} \bibinfo{person}{Felix~C. Freiling}.}
  \bibinfo{year}{2013}\natexlab{}.
\newblock \showarticletitle{Forensic Application-Fingerprinting Based on File
  System Metadata}. In \bibinfo{booktitle}{\emph{Seventh International
  Conference on {IT} Security Incident Management and {IT} Forensics, {IMF}
  2013, Nuremberg, Germany, March 12-14, 2013}},
  \bibfield{editor}{\bibinfo{person}{Holger Morgenstern}, \bibinfo{person}{Ralf
  Ehlert}, \bibinfo{person}{Felix~C. Freiling}, \bibinfo{person}{Sandra
  Frings}, \bibinfo{person}{Oliver G{\"{o}}bel}, \bibinfo{person}{Detlef
  G{\"{u}}nther}, \bibinfo{person}{Stefan Kiltz}, \bibinfo{person}{Jens Nedon},
  {and} \bibinfo{person}{Dirk Schadt}} (Eds.). \bibinfo{publisher}{{IEEE}
  Computer Society}, \bibinfo{pages}{98--112}.
\newblock
\urldef\tempurl%
\url{https://doi.org/10.1109/IMF.2013.20}
\showDOI{\tempurl}


\bibitem[Lazer et~al\mbox{.}(2026)]%
        {LazerAGB26}
\bibfield{author}{\bibinfo{person}{Sahaya~Jestus Lazer},
  \bibinfo{person}{Kshitiz Aryal}, \bibinfo{person}{Maanak Gupta}, {and}
  \bibinfo{person}{Elisa Bertino}.} \bibinfo{year}{2026}\natexlab{}.
\newblock \bibinfo{title}{A Survey of Agentic AI and Cybersecurity: Challenges,
  Opportunities and Use-case Prototypes}.
\newblock
\newblock
\showeprint[arxiv]{2601.05293}~[cs.CR]
\urldef\tempurl%
\url{https://arxiv.org/abs/2601.05293}
\showURL{%
\tempurl}


\bibitem[Meske et~al\mbox{.}(2025)]%
        {MeskeHWLB25}
\bibfield{author}{\bibinfo{person}{Christian Meske}, \bibinfo{person}{Tobias
  Hermanns}, \bibinfo{person}{Esther von~der Weiden},
  \bibinfo{person}{Kai{-}Uwe Loser}, {and} \bibinfo{person}{Thorsten Berger}.}
  \bibinfo{year}{2025}\natexlab{}.
\newblock \showarticletitle{Vibe Coding as a Reconfiguration of Intent
  Mediation in Software Development: Definition, Implications, and Research
  Agenda}.
\newblock \bibinfo{journal}{\emph{{IEEE} Access}}  \bibinfo{volume}{13}
  (\bibinfo{year}{2025}), \bibinfo{pages}{213242--213259}.
\newblock
\urldef\tempurl%
\url{https://doi.org/10.1109/ACCESS.2025.3645466}
\showDOI{\tempurl}


\bibitem[Metz(2025)]%
        {MetzArtifacts}
\bibfield{author}{\bibinfo{person}{Joachim Metz}.}
  \bibinfo{year}{2025}\natexlab{}.
\newblock \bibinfo{title}{Digital Forensics Artifacts Repository
  Documentation}.
\newblock
  \bibinfo{howpublished}{\url{https://github.com/ForensicArtifacts/artifacts}}.
\newblock
\urldef\tempurl%
\url{https://artifacts.readthedocs.io/en/latest/}
\showURL{%
\tempurl}
\newblock
\shownote{Accessed: 2026-02-03}.


\bibitem[Michelet et~al\mbox{.}(2026)]%
        {MicheletSWB26}
\bibfield{author}{\bibinfo{person}{Gaëtan Michelet}, \bibinfo{person}{Janine
  Schneider}, \bibinfo{person}{Aruna Withanage}, {and} \bibinfo{person}{Frank
  Breitinger}.} \bibinfo{year}{2026}\natexlab{}.
\newblock \showarticletitle{Hey GPT-OSS, looks like you got it – Now walk me
  through it! An assessment of the reasoning language models chain of thought
  process for digital forensics}.
\newblock \bibinfo{journal}{\emph{Forensic Science International: Digital
  Investigation}}  \bibinfo{volume}{56} (\bibinfo{year}{2026}),
  \bibinfo{pages}{302052}.
\newblock
\showISSN{2666-2817}
\urldef\tempurl%
\url{https://doi.org/10.1016/j.fsidi.2026.302052}
\showDOI{\tempurl}
\newblock
\shownote{DFRWS EU 2026 - Selected Papers from the 13th Annual Digital
  Forensics Research Conference Europe}.


\bibitem[Pati(2025)]%
        {Pati25}
\bibfield{author}{\bibinfo{person}{Ashis~Kumar Pati}.}
  \bibinfo{year}{2025}\natexlab{}.
\newblock \showarticletitle{Agentic {AI:} {A} Comprehensive Survey of
  Technologies, Applications, and Societal Implications}.
\newblock \bibinfo{journal}{\emph{{IEEE} Access}}  \bibinfo{volume}{13}
  (\bibinfo{year}{2025}), \bibinfo{pages}{151824--151837}.
\newblock
\urldef\tempurl%
\url{https://doi.org/10.1109/ACCESS.2025.3585609}
\showDOI{\tempurl}


\bibitem[Sapkota et~al\mbox{.}(2026)]%
        {SapkotaRK26}
\bibfield{author}{\bibinfo{person}{Ranjan Sapkota},
  \bibinfo{person}{Konstantinos~I. Roumeliotis}, {and} \bibinfo{person}{Manoj
  Karkee}.} \bibinfo{year}{2026}\natexlab{}.
\newblock \showarticletitle{{AI} Agents vs. Agentic {AI:} {A} Conceptual
  taxonomy, applications and challenges}.
\newblock \bibinfo{journal}{\emph{Inf. Fusion}}  \bibinfo{volume}{126}
  (\bibinfo{year}{2026}), \bibinfo{pages}{103599}.
\newblock
\urldef\tempurl%
\url{https://doi.org/10.1016/J.INFFUS.2025.103599}
\showDOI{\tempurl}


\bibitem[Scanlon et~al\mbox{.}(2023)]%
        {ScanlonBHHS23}
\bibfield{author}{\bibinfo{person}{Mark Scanlon}, \bibinfo{person}{Frank
  Breitinger}, \bibinfo{person}{Christopher Hargreaves},
  \bibinfo{person}{Jan-Niclas Hilgert}, {and} \bibinfo{person}{John Sheppard}.}
  \bibinfo{year}{2023}\natexlab{}.
\newblock \showarticletitle{ChatGPT for digital forensic investigation: The
  good, the bad, and the unknown}.
\newblock \bibinfo{journal}{\emph{Forensic Science International: Digital
  Investigation}}  \bibinfo{volume}{46} (\bibinfo{year}{2023}),
  \bibinfo{pages}{301609}.
\newblock
\showISSN{2666-2817}
\urldef\tempurl%
\url{https://doi.org/10.1016/j.fsidi.2023.301609}
\showDOI{\tempurl}


\bibitem[Schneider and Breitinger(2023)]%
        {SchneiderB23}
\bibfield{author}{\bibinfo{person}{Johannes Schneider} {and}
  \bibinfo{person}{Frank Breitinger}.} \bibinfo{year}{2023}\natexlab{}.
\newblock \showarticletitle{Towards {AI} forensics: Did the artificial
  intelligence system do it?}
\newblock \bibinfo{journal}{\emph{J. Inf. Secur. Appl.}}  \bibinfo{volume}{76}
  (\bibinfo{year}{2023}), \bibinfo{pages}{103517}.
\newblock
\urldef\tempurl%
\url{https://doi.org/10.1016/J.JISA.2023.103517}
\showDOI{\tempurl}


\bibitem[Shapira et~al\mbox{.}(2026)]%
        {ShapiraWYSPFBLJPC26}
\bibfield{author}{\bibinfo{person}{Natalie Shapira}, \bibinfo{person}{Chris
  Wendler}, \bibinfo{person}{Avery Yen}, \bibinfo{person}{Gabriele Sarti},
  \bibinfo{person}{Koyena Pal}, \bibinfo{person}{Olivia Floody},
  \bibinfo{person}{Adam Belfki}, \bibinfo{person}{Alexander~R. Loftus},
  \bibinfo{person}{Aditya~Ratan Jannali}, \bibinfo{person}{Nikhil Prakash},
  \bibinfo{person}{Jasmine Cui}, \bibinfo{person}{Giordano Rogers},
  \bibinfo{person}{Jannik Brinkmann}, \bibinfo{person}{Can Rager},
  \bibinfo{person}{Amir Zur}, \bibinfo{person}{Michael Ripa},
  \bibinfo{person}{Aruna Sankaranarayanan}, \bibinfo{person}{David Atkinson},
  \bibinfo{person}{Rohit Gandikota}, \bibinfo{person}{Jaden Fiotto{-}Kaufman},
  \bibinfo{person}{EunJeong Hwang}, \bibinfo{person}{Hadas Orgad},
  \bibinfo{person}{P.~Sam Sahil}, \bibinfo{person}{Negev Taglicht},
  \bibinfo{person}{Tomer Shabtay}, \bibinfo{person}{Atai Ambus},
  \bibinfo{person}{Nitay Alon}, \bibinfo{person}{Shiri Oron},
  \bibinfo{person}{Ayelet Gordon{-}Tapiero}, \bibinfo{person}{Yotam Kaplan},
  \bibinfo{person}{Vered Shwartz}, \bibinfo{person}{Tamar~Rott Shaham},
  \bibinfo{person}{Christoph Riedl}, \bibinfo{person}{Reuth Mirsky},
  \bibinfo{person}{Maarten Sap}, \bibinfo{person}{David Manheim},
  \bibinfo{person}{Tomer Ullman}, {and} \bibinfo{person}{David Bau}.}
  \bibinfo{year}{2026}\natexlab{}.
\newblock \showarticletitle{Agents of Chaos}.
\newblock \bibinfo{journal}{\emph{CoRR}}  \bibinfo{volume}{abs/2602.20021}
  (\bibinfo{year}{2026}).
\newblock
\urldef\tempurl%
\url{https://doi.org/10.48550/ARXIV.2602.20021}
\showDOI{\tempurl}
\showeprint[arXiv]{2602.20021}


\bibitem[Spichiger and Adelstein(2025)]%
        {SpichigerA25}
\bibfield{author}{\bibinfo{person}{Hannes Spichiger} {and}
  \bibinfo{person}{Frank Adelstein}.} \bibinfo{year}{2025}\natexlab{}.
\newblock \showarticletitle{Preserving meaning of evidence from evolving
  systems}.
\newblock \bibinfo{journal}{\emph{Digit. Investig.}} \bibinfo{volume}{52},
  \bibinfo{number}{Supplement} (\bibinfo{year}{2025}), \bibinfo{pages}{301867}.
\newblock
\urldef\tempurl%
\url{https://doi.org/10.1016/J.FSIDI.2025.301867}
\showDOI{\tempurl}


\bibitem[Walker et~al\mbox{.}(2024)]%
        {WalkerGAHB24}
\bibfield{author}{\bibinfo{person}{Clinton Walker}, \bibinfo{person}{Taha
  Gharaibeh}, \bibinfo{person}{Ruba Alsmadi}, \bibinfo{person}{Cory~Lloyd
  Hall}, {and} \bibinfo{person}{Ibrahim~M. Baggili}.}
  \bibinfo{year}{2024}\natexlab{}.
\newblock \showarticletitle{Forensic Analysis of Artifacts from Microsoft's
  Multi-Agent {LLM} Platform AutoGen}. In \bibinfo{booktitle}{\emph{Proceedings
  of the 19th International Conference on Availability, Reliability and
  Security, {ARES} 2024, Vienna, Austria, 30 July 2024 - 2 August 2024}}.
  \bibinfo{publisher}{{ACM}}, \bibinfo{pages}{198:1--198:9}.
\newblock
\urldef\tempurl%
\url{https://doi.org/10.1145/3664476.3670908}
\showDOI{\tempurl}


\bibitem[Wickramasekara and Scanlon(2024)]%
        {WickramasekaraS24}
\bibfield{author}{\bibinfo{person}{Akila Wickramasekara} {and}
  \bibinfo{person}{Mark Scanlon}.} \bibinfo{year}{2024}\natexlab{}.
\newblock \showarticletitle{A Framework for Integrated Digital Forensic
  Investigation Employing AutoGen AI Agents}. In \bibinfo{booktitle}{\emph{2024
  12th International Symposium on Digital Forensics and Security (ISDFS)}}.
  \bibinfo{pages}{01--06}.
\newblock
\urldef\tempurl%
\url{https://doi.org/10.1109/ISDFS60797.2024.10527235}
\showDOI{\tempurl}


\end{thebibliography}

\appendix

\section{Supplementary Material}
\label{appendix:supplementary_material}
We provide a supplementary repository alongside this paper, containing the machine-readable description of relevant forensic artifacts, a Python-based tool to support the forensic analysis of OpenClaw, and documentation of the analysis process in the form of the \texttt{DFXML} outputs for our experiments.

\subsection{Artifact Examiner Tool}

To facilitate the analysis of OpenClaw, we developed the \textbf{\texttt{artifact-examiner}}, a Python-based command-line interface (CLI) tool included in the \texttt{artifact-examiner/} directory.
This tool leverages our analysis results to automatically ingest, parse, and correlate forensic data from an OpenClaw installation.

The Artifact Examiner offers several capabilities for investigators:
\begin{itemize}
[  leftmargin=1.5em,
    itemsep=-1ex,
    topsep=1ex,
    partopsep=0ex,
    parsep=1ex,
    labelsep=0.5em,]
    \item \textbf{Unified Timeline Analysis}: It reconstructs a chronological view of all agent activities by merging events from logs, session transcripts, and configuration changes. This allows analysts to trace the sequence of actions leading up to a specific event.
    \item \textbf{Session Inspection}: The tool provides interactive browsers for exploring agent sessions, including the ability to reconstruct conversation flows, tool invocations, and subagent spawns.
    \item \textbf{Anti-Forensics Detection}: By comparing execution logs with persistent session records, the tool can identify discrepancies that may indicate deleted sessions or tampered data, highlighting potential anti-forensic activities.
    \item \textbf{Capability Analysis}: It tracks the evolution of the agent's capabilities over time, such as changes in available tools, models, and permissions, providing insight into the agent's potential impact on the system.
\end{itemize}
Refer to the repository for a comprehensive listing of its capabilities. During our own analyses, we noted that the \texttt{artifact-examiner} reduces the manual effort required to analyze human interaction and agent behavior, while ensuring a consistent and reproducible investigation process.

\subsection{Forensic Artifact Definition}

We provide a technical and concise description of forensic artifacts via the \texttt{artifacts/data/openclaw.yml} file following the specification of the \emph{Digital Forensics Artifacts Repository}~\citep{MetzArtifacts}.
This YAML file serves as a machine-readable schema that defines the location, structure, and semantic meaning of important artifacts generated by the OpenClaw agent. Each artifact entry includes:
\begin{itemize}
[  leftmargin=1.5em,
    itemsep=-1ex,
    topsep=1ex,
    partopsep=0ex,
    parsep=1ex,
    labelsep=0.5em,]
    \item \sloppy \textbf{Name}: A unique identifier for the artifact (e.g., \texttt{Linux\-OpenClaw\-Session\-Transcript\-Jsonl}, \texttt{Linux\-OpenClaw\-Daily\-Json\-LogFiles}).
    \item \textbf{Documentation}: A brief description of the artifact's purpose, content, and forensic significance. This documentation is embedded directly within the schema to ensure it remains synchronized with the implementation.
    \item \textbf{Sources}: The file paths or glob patterns where the artifact can be found on the target system (e.g., \texttt{\textasciitilde/.openclaw/agents/<agentId>/sessions/*.jsonl}).
\end{itemize}

We aim to enable both human analysts and automated tools to reliably locate and interpret forensic data of OpenClaw.

\subsection{Experimental Methodology}
The artifact documentation within the \texttt{artifacts/} directory was informed through a controlled series of experiments designed to isolate file system changes associated with specific user and agent actions.
The methodology involved the following steps:

\begin{enumerate}
[  leftmargin=1.5em,
    itemsep=-1ex,
    topsep=1ex,
    partopsep=0ex,
    parsep=1ex,
    labelsep=0.5em,]
    \item \textbf{Baseline Establishment}: A clean virtual machine (VM) snapshot was created with the base operating system and necessary dependencies but without the OpenClaw agent installed.
    \item \textbf{Action Execution}: Specific actions were performed in a sequential manner, such as installing the agent, onboarding, sending text and voice messages, adding calendar events, and spawning subagents.
    \item \textbf{Differential Analysis}: After each action, the VM state was captured, and a differential analysis was performed against the previous state to identify created, modified, or deleted files.
\end{enumerate}

\subsection{Availability}

The supplementary material, including the analysis tool \texttt{artifact-examiner}, the \texttt{openclaw.yml} schema, and the differential analysis, is published at:

\medskip

\noindent\url{\Repo}

\medskip

\noindent Artifacts have been redacted where necessary to remove sensitive personal information.

\end{document}